\documentclass[twocolumn,pra,a4paper,showpacs,floatfix,preprintnumbers,amsmath,amssymb,superscriptaddress]{revtex4}

\usepackage{amsmath}
\usepackage{latexsym}
\usepackage{graphicx}
\usepackage{amssymb}

\usepackage{graphicx}
\usepackage{dcolumn}
\usepackage{bm}

\newcommand{\ud}{\mathrm{d}}
\newcommand{\be}{\begin{equation}}
\newcommand{\ee}{\end{equation}}
\newcommand{\beq}{\begin{eqnarray}}
\newcommand{\eeq}{\end{eqnarray}}
\newcommand{\out}{\mathrm{out}}
\newcommand{\inp}{\mathrm{inp}}
\newcommand{\Tr}{\mathrm{Tr}}
\newcommand{\QND}{\mathrm{QND}}

\begin{document}

\title{Ultrasensitive Atomic clock with single-mode number-squeezing}

\author{L. Pezz\'e}\affiliation{
Laboratoire Charles Fabry de l'Institut d'Optique,
campus Polytechnique RD128, F-91127 Palaiseau cedex, France}
\author{A. Smerzi}\affiliation{ CNR-INFM BEC center and Dipartimento di Fisica, Universit\`a di Trento, I-38050 Povo, Italy}

\date{\today}
                      
\begin{abstract}
We show that the sensitivity of an atomic clock
can be enhanced 
below the shot-noise level by 
initially squeezing, and then measuring in output, the population 
of a \emph{single} atomic level. 
This can simplify current experimental protocols which requires  
squeezing of the relative number of particles of the two populated states.
We finally study, as a specific application, the clock sensitivity obtained with a
single mode quantum non-demolition measurement. 
\end{abstract}


\maketitle

{\it Introduction}. 
Atomic clocks and interferometers are among the most sensitive measurement
devices available within the current technology \cite{Cronin_2008}.
Their precision is bounded by the fundamental noise
imposed by quantum mechanics uncertainties.  
With uncorrelated atoms, the phase sensitivity of the Ramsey sequence scales as the inverse 
square root of the total number of particles, the so-called 
shot noise (or standard quantum) limit, first observed 
experimentally in \cite{Santerelli_1999}.
The possibility to overcome the shot noise limit by quantum engineering 
specific atomic correlations is a break grounding prediction which is under intense experimental investigation. 
Most current efforts focus on the creation of spin-squeezed states \cite{Gross_2010, 
Appel_2009, Schleier_2009, Esteve_2008, Riedel_2010}. 
This can be achieved, for instance, by manipulating a cloud of cold atoms 
via the back-reaction of quantum non-demolition (QND) measurements  
\cite{Appel_2009, Schleier_2009} or with interaction-induced nonlinearity 
using Bose Einstein Condensates (BECs) \cite{Gross_2010, Riedel_2010}.
With BEC, entangled-enhanced Ramsey phase sensitivity has been recently 
demonstrated \cite{Gross_2010}.

In the current literature, it is shown that sub shot noise phase sensitivity
in an atomic clock is generally associated to spin squeezing.
In this manuscript we demonstrate a sub shot noise phase sensitivity up to the Heisenberg limit 
with the initial squeezing and the measurement in output of the particle population 
of a \emph{single} atomic level. This can simplify current interferometric protocols. 
A non-destructive atom-light interaction of a single clock level has been experimentally 
demonstrated in \cite{Windpassinger_2008, Chaudhury_2006}. 
Therefore, our prediction can be readily tested experimentally and find application 
for precision atomic sensors within the present state-of-the-art technology.

{\it Sub shot-noise with single mode squeezing}.
We consider two atomic clock levels 
$a$ and $b$ of energy $\hbar \omega_a$ and $\hbar \omega_b$, respectively.
The goal is to 
estimate the frequency difference $\Delta \omega = \omega_a - \omega_b$ 
with the highest possible sensitivity.
The Ramsey interferometric sequence consists of four steps:
a Rabi $\pi/2$ pulse of constant power and frequency $\omega$  
(as close as possible in resonance with the atomic transition)
applied for a time $\tau = \pi/2 \Omega_R$, being $\Omega_R$ the Rabi frequency.
Then the system freely evolves 
for a period of time $T$. Finally, after a second $\pi/2$ pulse, 
the number of particles is measured in a single output level.
The quantum mechanical expectation value of the number of particles  
in the output $a$ mode is given by 
$\langle \hat{n}_a \rangle_{\mathrm{out}}=
\langle \hat{n}_a \rangle_{\mathrm{inp}} \cos^2(\frac{\theta}{2}) + 
\langle \hat{n}_b \rangle_{\mathrm{inp}} \sin^2(\frac{\theta}{2}) -
\frac{1}{2} \langle \hat{a}^{\dag} \hat{b} + \hat{b}^{\dag} \hat{a} \rangle_{\mathrm{inp}} \sin \theta$ \cite{nota2},
where $\hat{a}$ ($\hat{a}^{\dag}$) and $\hat{b}$ ($\hat{b}^{\dag}$) 
are particle annihilation (creation) operators of the 
$a$ and $b$ mode, respectively, and 
$\hat{n}_a \equiv \hat{a}^{\dag} \hat{a}$ ($\hat{n}_b \equiv \hat{b}^{\dag} \hat{b}$)
is the number of particles operator.
The quantity $\langle \hat{n}_a \rangle_{\mathrm{out}}$ 
depends on the phase $\theta = \delta \times T$ accumulated 
during the free precession, being 
$\delta= \Delta \omega - \omega$ the detuning of the Rabi pulse from the atomic transition.
By collecting $m$ measurements with results $n_{a}^{(1)},...,n_{a}^{(m)}$, we can calculate the average 
number of particles in the $a$ output mode $\bar n_{a}^{out} = \sum_{i=1}^m n_{a}^{(i)} /m$.
The phase is inferred by approximating the expectation value 
$\langle \hat{a}^{\dag} \hat{a} \rangle_{\mathrm{out}}$ with the classical average 
$\bar n_{a}^{\mathrm{out}}$ and inverting the equation
$\bar n_{a}^{\mathrm{out}}=\langle \hat{a}^{\dag} \hat{a} \rangle_{\mathrm{inp}} \cos^2(\frac{\theta_{\mathrm{est}}}{2}) + 
\langle \hat{b}^{\dag} \hat{b} \rangle_{\mathrm{inp}} \sin^2(\frac{\theta_{\mathrm{est}}}{2}) -
\frac{1}{2} \langle \hat{a}^{\dag} \hat{b} + \hat{b}^{\dag} \hat{a} \rangle_{\mathrm{inp}} \sin \theta_{\mathrm{est}}$,
where $\theta_{\mathrm{est}}$ is the estimated value of $\theta$.
In the central limit, the expected phase sensitivity is
\be \label{err}
\Delta \theta = \frac{(\Delta \hat n_{a})_{\out}}{ \sqrt{m} \, 
\vert\ud \langle \hat n_{a} \rangle_{\out} /\ud \theta \vert} 
\bigg\vert_{\theta =\theta_{\mathrm{est}}},
\ee 
where $\Delta \theta$ is the variance of $\theta_{\mathrm{est}}$, 
calculated with error propagation, assuming that $\bar n_{a}^{\out}$ fluctuates with 
variance $(\Delta \hat n_{a} )_{\out}^2 = 
\langle \hat n_{a}^2 \rangle_{\out} - \langle \hat n_{a} \rangle_{\out}^2$.
In particular, when $\theta \sim 0$ and with initial symmetric populations 
$\langle \hat n_{a} \rangle_{\inp} = \langle \hat n_{b} \rangle_{\inp}$,
Eq.(\ref{err}) becomes
\be \label{err1}
\Delta \theta = \frac{2 (\Delta \hat n_{a} )_{\inp}}
{\sqrt{m} \, |\langle \hat{a}^{\dag} \hat{b} + \hat{b}^{\dag} \hat{a} \rangle_{\inp}|}
\bigg\vert_{\theta =\theta_{\mathrm{est}}}.
\ee
Equations (\ref{err}) and (\ref{err1})
relate the interferometric phase sensitivity $\Delta \theta$ to the fluctuations of the number of particles in a single input atomic level,
and capture the main results of this manuscript.
In particular, according to Eq.(\ref{err1}), sub shot-noise phase sensitivity 
can be obtained by squeezing the population fluctuations in the input $a$, 
$(\Delta \hat n_{a} )_{\inp}^2 < \langle \hat{n} \rangle/2$,
while keeping the coherence between the two mode 
$\langle \hat{a}^{\dag} \hat{b} \rangle_{\inp} \sim \langle \hat{n} \rangle/2$. 
Here $\langle \hat{n} \rangle = \langle \hat n_{a} \rangle_{\inp} + \langle \hat n_{b} \rangle_{\inp}$
is the average number of particles in the input state.
As an example, we consider the product of two Gaussian pure states 
$|\psi\rangle_{a,b} \propto  \sum_{n=0}^{+\infty} 
\exp[{-(n-\langle \hat{n} \rangle/2)^2/4 \sigma_{a,b}^2}] |n\rangle_{a,b}$
as input of the interferometer,
where $\sigma_b = \sqrt{\langle \hat{n} \rangle/2}$ and $\sigma_a = \kappa \sqrt{\langle \hat{n} \rangle/2}$.
Number-squeezing in the $a$ mode \cite{nota0} is obtained for $\kappa <1$, and 
equation (\ref{err1}) predicts a sub shot noise sensitivity 
$\Delta \theta = \kappa /\sqrt{m \langle \hat{n} \rangle}$.
In the Fock limit $\sigma_a \to 0$, Eq.(\ref{err1}) predicts the ultimate Heisenberg scaling
\be \label{focklimit}
\Delta \theta = \frac{\sqrt{2}}{\langle \hat{n} \rangle \sqrt{m}}.
\ee  
The Fock state limit is particularly interesting and will be further discussed below.

Notice that Eq. (\ref{err1}) resembles the familiar relation between interferometric sensitivity and 
spin squeezing \cite{Kitagawa_1993, Wineland_1994, Kuzmich_1998}. 
However, since the total number of particles is not fixed, 
number squeezing in a single clock level does not necessarily 
implies the squeezing of the relative population between the two input modes.
Only when the total number of particles is fixed, the two become equivalent.
Our analysis includes this special case.

{\it Entanglement and Fock state limit.}
The relation between multiparticle-entaglement and 
sub shot noise phase sensitivity in a linear interferometer 
has been recently discussed for states of fixed \cite{Pezze_2009} 
and fluctuating \cite{Hyllus} number of particles (qubits).
In particular, states satisfying the inequality 
$F_Q[\hat{\rho}, \hat{J}_y]> \langle \hat{n} \rangle$, being $F_Q[\hat{\rho}, \hat{J}_y]$ 
the quantum Fisher information \cite{Pezze_2009,Hyllus} and 
$\hat{J}_y=(\hat{a}^{\dag} \hat{b} + \hat{b}^{\dag} \hat{a})/2$ \cite{note_Schwinger}, 
are entangled and provide a sub shot-noise phase sensitivity in a Ramsey interferometer.
The optimal phase sensitivity, after $m$ independent measurements, is
$\Delta \theta = 1/\sqrt{m F_Q[\hat{\rho}, \hat{J}_y]}$.
In particular, for an input state 
\be \label{Fock}
\hat{\rho}_{\mathrm{inp}}=|N\rangle_a \langle N| \otimes \hat{\rho}_b,
\ee
being $\hat{\rho}_b = \sum_n \rho_n \vert n \rangle_b \langle n \vert$ a generic density matrix of 
$\langle \hat n_{b} \rangle_{\inp} =\sum_{n=0}^{+\infty} \rho_n n$ average number of particles,
we find
\be \label{CRLB}
\Delta \theta = \frac{1}{\sqrt{m} \sqrt{2N \langle \hat n_{b} \rangle_{\inp} +N+ \langle \hat n_{b} \rangle_{\inp} }}.
\ee
Equation (\ref{CRLB}) is characterized by interesting limits.
If the mode $b$ is left empty, $\rho_n=\delta_{n,0}$,  
the sensitivity of the clock is given by the standard quantum limit (SQL), 
$\Delta \theta = \frac{1}{\sqrt{ m \langle \hat{n} \rangle}}$,
where $\langle \hat{n} \rangle = N + \langle \hat n_{b} \rangle_{\inp}$ 
is the average number of particles in the input state
(in this case $\langle \hat{n} \rangle=N$).
Conversely, if the mode $b$ is left almost (but not completely)
empty, with $\langle \hat n_{b} \rangle_{\inp} \ll N$,
the sensitivity $\Delta \theta \approx 1 /  \sqrt{2 \langle \hat n_{b} \rangle_{\inp} + 1} \sqrt{m \langle \hat{n} \rangle}$
is below the SQL by a factor $\sqrt{2 \langle \hat n_{b} \rangle_{\inp} + 1}$.
Moreover, at the optimal condition $N = \langle \hat n_{b} \rangle_{\inp}$, Eq.(\ref{CRLB}) 
predicts $\Delta \theta \approx \frac{\sqrt{2}}{\langle \hat{n} \rangle \sqrt{m}}$, 
independently of the input states $\hat{\rho}_b$, and, in particular, we 
recover Eq.(\ref{focklimit}).
Measuring the number of particles in a single output port not only provides
an optimal estimation strategy but also saturates the Heisenberg limit \cite{Hyllus}. 
As anticipated above, we note that Eq.(\ref{Fock}) is not a spin squeezed state.  
The spin squeezing parameter 
$\xi^2=(\Delta \hat{J}_z)_{\inp}^2/ (\langle \hat{J}_x\rangle_{\inp}^2  + \langle \hat{J}_y\rangle_{\inp}^2)$
\cite{Wineland_1994, Sorensen_2001} diverges in general when calculated for the input state Eq.(\ref{Fock}), 
and is undetermined, $\xi^2=0/0$, when also the input $b$ is in a Fock state ($\rho_n=\delta_{n,N}$) \cite{Kim_1998}. 
In the latter case, corresponding to the twin Fock state,  
Eq.(\ref{CRLB}) recovers the Heisenberg limit sensitivity first predicted in \cite{Holland_1993}.

\begin{figure}[!t]
\begin{center}
\includegraphics[scale=0.31]{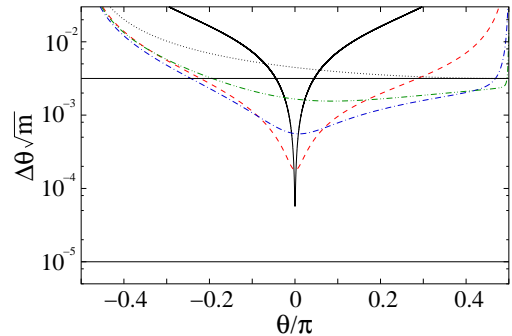}
\end{center}
\caption{\small{Phase sensitivity Eq.(\ref{err}) as function of $\theta$ and for different values of $\gamma$:
for $\gamma/\pi=0$ (dotted black line) the sensitivity is given by Eq.(\ref{Dtheta1}),  
$\gamma/\pi=10^{-5}$ (dot-dot-dashed green line), 
$\gamma/\pi=10^{-4}$ (dot-dashed blue line), 
$\gamma/\pi=10^{-3}$ (dashed red line), 
$\gamma/\pi=10^{-2}$ (solid black line). 
The curves have been obtained for a Gaussian distribution of the total number of particles ($\sigma^2=\langle\hat{n}\rangle$)
and $\langle\hat{n}\rangle=10^5$ atoms. 
The upper (lower) horizontal lines is 
$1/\sqrt{\langle\hat{n}\rangle}$ ($\sqrt{2}/\langle\hat{n}\rangle$).}} \label{figure1} 
\end{figure}

\begin{figure}[!t]
\begin{center}
\includegraphics[scale=0.26]{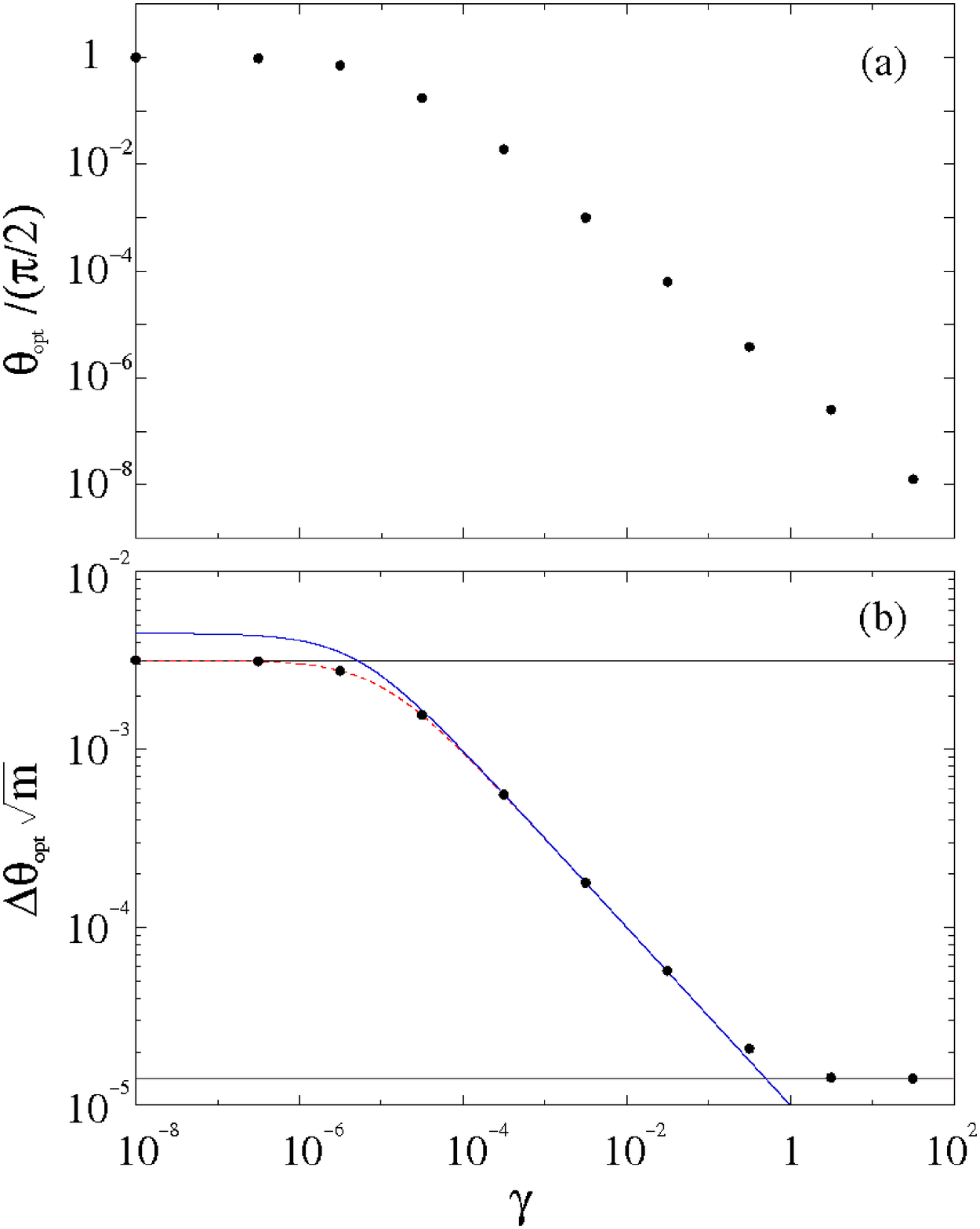}
\end{center}
\caption{\small{a) Optimal phase shift $\theta_{opt}$ (minimum of Eq.(\ref{err})) as function of $\gamma$.
b) Phase sensitivity Eq.(\ref{err}) calculated at the optimal point $\theta_{opt} $as function of $\gamma$ (dots).
The lines are Eq.(\ref{Dtheta}) for $\sigma^2=\langle\hat{n}\rangle$ (blue line), and $\sigma=0$ (dotted red line).
The horizontal lines are $1/\sqrt{\langle\hat{n}\rangle}$ (upper line) and  
Eq.(\ref{focklimit}) (lower line).}} \label{figure2} 
\end{figure}

\emph{QND state preparation.}
We now discuss a possible experimental implementation of the protocol discussed above.
The single-mode number squeezing is produced by a
QND interaction between the atoms cloud and a light field, recently demonstrated in 
\cite{Windpassinger_2008, Chaudhury_2006}.
The initial atomic cloud, obtained by optically pumping the atoms to a single level, 
is described by the density matrix  
$\hat{\rho} = \sum_{N=0}^{+ \infty} P_N  \, |N,0 \rangle  \langle N, 0 |$,
where $|N,0 \rangle  \equiv |N\rangle_a |0\rangle_b$ 
is the state with $N$ atoms in the levels $a$ and 0 atoms in level $b$, which occurs 
with probability $P_N$ ($P_N >0$ and $\sum_N P_N =1$). 
In order to equally populate in average the two input modes, we apply a $\pi/2$ pulse. 
The density matrix becomes
$\hat{\rho}_A^{(0)} = \sum_{N=0}^{+ \infty} P_{N} |\psi_N\rangle \langle \psi_N |$
where 
$|\psi_N\rangle = \sum_{n=0}^{N} \frac{1}{2^{N}} \sqrt{\frac{N!}{n! (N-n)!}} \big|n, N-n \big\rangle$.
The number of particles distribution in each mode has
$\langle \hat{n}_a \rangle = \langle \hat{n}_b \rangle =  \langle\hat{n}\rangle/2$, 
$(\Delta \hat{n}_a)^2 = (\Delta \hat{n}_b)^2 = (\sigma^2 +\langle\hat{n}\rangle)/4$
where $\langle\hat{n}\rangle = \sum_N P_N N $ and $\sigma^2 = \sum_N P_N (N-\langle\hat{n}\rangle)^2$.
With this state as input of the clock, Eq.(\ref{err}) predicts 
\be \label{Dtheta1}
\Delta \theta=\frac{1}{\sqrt{m}}\sqrt{\frac{1}{\langle\hat{n}\rangle}+\frac{\sigma^2}{\langle\hat{n}\rangle^2}\frac{(1-\sin \theta )^2}{\cos^2 \theta }}.
\ee
At the optimal value of the phase shift $\theta=\pi/2$, the phase sensitivity is at 
the standard quantum limit $\Delta \theta=1/\sqrt{m \langle\hat{n}\rangle}$.
It is possible to overcome this limit by squeezing the number of particles fluctuations in one mode. 
This is done here by letting the atoms in $a$ to interact with a coherent light field $|\alpha \rangle$ 
(we take, without loss of generality, the amplitude $\alpha$ to be real) via the QND Hamiltonian 
\be \label{H}
\hat{H}_{\QND} = \hbar g  \, (\hat{c}^{\dag} \hat{c}) \, (\hat{a}^{\dag} \hat{a}),
\ee
where $\hat{c}^{\dag}$ ($\hat{c}$) created (annihilate) a photon of light field mode and
$g$ is the coupling parameter. 
The Hamiltonian Eq.(\ref{H}) describes an out of resonance interaction of the light field 
with the atoms in the level $a$ \cite{Kuzmich_1998}. We assume here that the detuning of the 
light is large enough to adiabatically eliminate the 
excited state population and neglect 
possible decoherence effects \cite{Bouchoule_2002}. 
The coupling strength is normally weak but can be enhanced by placing 
the atoms inside an optical cavity \cite{Nielsen_2008}.
The entangled atom-light system is described by the density matrix 
$\hat{\rho}_{AL}^{(0)}=e^{i \hat{H}_{\QND} t/\hbar} \hat{\rho}_A^{(0)} \otimes |\alpha \rangle \langle \alpha | e^{-i \hat{H}_{\QND} t/\hbar}$, 
where $t$ is the interaction time between the light and the atomic cloud and 
$\alpha$ is the amplitude of the coherent state. 
The effect of the QND interaction is to shift the phase of the light by
a quantity proportional to the number of atoms.
There are different experimental possibilities to estimate the phase shift \cite{Windpassinger_2008, Chaudhury_2006}.
In the following we consider the homodyne measurement of the 
$p$ quadrature, $\hat{p} = (\hat{c} - \hat{c}^{\dag})/2i$. 
The probability to measure the eigenvalue $p$ is given by 
$P_0(p) = \mathrm{Tr}\Big[\hat{\rho}_{L}^{(0)} |p\rangle \langle p |\Big]$,
where $\hat{\rho}_{L}^{(0)} =  \mathrm{Tr}_A[\hat{\rho}_{AL}^{(0)}]$ is traced over the atomic degrees of freedom \cite{nota3}. 
After the measurement, with result $p_0$,
the density matrix of the atomic cloud becomes
$\hat{\rho}_A^{(1)} = \mathrm{Tr}_L \Big[\hat{\rho}_{AL}^{(0)} |p_0 \rangle \langle p_0 |\Big] / P_0(p_0)$.
The back reaction effect induced by the measurement \cite{Braginsky_book} reduces the fluctuations 
of the number of atoms in the $a$ mode.
In the limit $\Omega \langle\hat{n}\rangle \ll 1$, with $\Omega=gt$, the quadrature probability is a sum of Gaussian functions \cite{nota4} and we can distinguish
two regimes.
When $\alpha \Omega \gg 1$, the measurement result is only compatible with a precise number of atoms and 
therefore the QND interaction projects the atomic population of the $a$ mode to a Fock state. In this case the sensitivity
is at the Heisenberg limit Eq.(\ref{focklimit}).
Conversely, when $\alpha \Omega \lesssim 1$, which is the most realistic scenario from the experimental point of view,
the result of the quadrature measurement is compatible with several values of 
the number of particles in the $a$ mode, and the back reaction only produces a moderate squeezing. 
Therefore, in order to increase the squeezing, it is necessary to repeat $M\gg 1$ times the QND protocol \cite{notaM}.
Analytical results can be obtained in the limit $\gamma = \alpha^2 \Omega^2 M \lesssim 1$.
In this case we obtain
$(\Delta \hat{a}^{\dag} \hat{a})^2 = \frac{(\sigma^2 +\langle\hat{n}\rangle)/4}{1+ \gamma (\sigma^2 +\langle\hat{n}\rangle)}$ and
$\langle \hat{a}^{\dag} \hat{b} + \hat{b}^{\dag} \hat{a} \rangle_{\inp} \approx 2 \langle\hat{n}\rangle \cos(M \alpha^2 \Omega)$.
Conversely, the input $b$ retains, in average, the initial number fluctuations.
By tuning $M \alpha^2 g$ to values close to an integer multiple of $\pi$, 
Eq.(\ref{err1}) gives 
\be \label{Dtheta}
\Delta \theta = \frac{1}{\langle\hat{n}\rangle\sqrt{m}}\sqrt{ \frac{\sigma^2 +\langle\hat{n}\rangle}{1+ \gamma (\sigma^2 +\langle\hat{n}\rangle)}}.
\ee
Equation (\ref{Dtheta}) is obtained for $\theta \sim 0$ and arbitrary fluctuations $\sigma$.
In particular, when $\gamma > \sigma^2/ \bar n (\bar n + \sigma^2)$, 
we obtain a sub shot noise ($\Delta \theta < 1/\sqrt{\langle\hat{n}\rangle m}$) limit.
In figure (\ref{figure1}) we plot the phase sensitivity, calculated using Eq.(\ref{err}) with a coherent state having $\sigma^2 = \langle\hat{n}\rangle = 10^5$,
as a function of the phase shift $\theta$. The different lines correspond to different values of $\gamma$.
In particular, the dotted black line is obtained for $\gamma=0$ and is given by Eq.(\ref{Dtheta1}).
The upper (lower) horizontal lines is $1/\sqrt{\langle\hat{n}\rangle}$ ($\sqrt{2}/\langle\hat{n}\rangle$).
The figure shows that sub shot noise can be obtained around an optimal value of the phase shift, $\theta_{opt}$ (minimum of each curve).
In figure (\ref{figure2},a) we show $\theta_{opt}$ as a function of $\gamma$: it is close to 
$\theta_{opt} \sim \pi/2$ for moderate squeezing and rapidly tends to $\theta_{opt} \sim 0$ by increasing $\gamma$.
In figure (\ref{figure2},b) we show the optimal phase sensitivity
as a function of $\gamma$. Numerical results (circles) agree with the solid blue line 
given by Eq.(\ref{Dtheta}) in the limit of relatively small $\gamma$, 
where the optimal phase shift is $\theta_{opt} \sim 0$ \cite{notanotice}.
For $\gamma > 1$ the QND project the state close to Fock, 
and the phase sensitivity converges to Eq.(\ref{focklimit}) (lower solid horizontal line).
Notice that Eq.(\ref{Dtheta}) applies also when the total number of particles 
is fixed ($\sigma=0$). In this case, the squeezing of the number of particles of a single 
level is equivalent to relative number squeezing. 
In figure (\ref{figure2},B) the dotted red line corresponding to the case 
$\sigma=0$ (and $\langle\hat{n}\rangle=10^5$ atoms),
superposes to the dots, obtained from numerical simulations with $\sigma^2=\langle\hat{n}\rangle$.
This clearly shows that relative number squeezing does not provide any advantage with respect to  
single mode number squeezing: for the same number of particles and squeezing parameters
we obtain the same level of sub shot noise interferometric sensitivity.

\emph{Discussion.} 
As discussed above, a QND measurement on a {\it single} energy level is sufficient to 
prepare an input states useful to reach a sub shot noise phase sensitivity. This 
can potentially simplify current experimental schemes. 
The reduction of the fluctuations of the relative 
population (while preserving the coherence) with a QND atom-light interaction 
has been recently demonstrated. This has been done by i) 
addressing the two atomic levels with two carefully detuned laser beams \cite{Appel_2009}
or ii) with a single laser beam tuned at a very precise frequency and detuning 
where the index of refraction of the gas crosses a zero value \cite{Schleier_2009}.
On the other hand, when addressing a single level, it is possible to perform QND with the freedom of
choosing the laser frequency and detuning so to minimize the decoherence effect due to 
spontaneous emission. 
As shown in \cite{Appel_2009, Schleier_2009} this is one of the main
limitations of the level of squeezing reached experimentally.
Within our scheme, it is also possible to select the atomic transition 
in order to maximize the response of the light field to the atom-light interaction.
A further experimental advantage of our proposal is that the clock frequency can be estimated by measuring
the number of particles in a single output port of the Ramsey interferometer.
This avoids the further noise introduced by inverting and measuring the population
of the second mode, as currently done.
We thus expect that our protocol might experimentally provide higher squeezing 
with a more robust apparatus.

\emph{Conclusion.} 
We have discussed a new protocol to reach a sub shot noise sensitivity (up to the Heisenberg limit) in an atomic clock.
This requires: i) reduced particles number fluctuations in a single input mode and 
ii) the measurement of the number of particles in a single output.
Our results can be interpreted in terms of useful entanglement created 
by squeezing the population fluctuations in a single mode. 
We provide a simple analysis of the number squeezing produced by a QND interaction between a light field and 
the atomic sample showing, for this experimentally relevant situation, the possibility to readily verify our predictions. 
Since Fock states of a small number of photons are currently experimentally available \cite{Ourjoumtsev_2007}, 
our predictions can be relevant also in the optical domine.

\emph{Acknowledgement.}
We thank M. Barbieri, A. Bertoldi, C.L. Garrido-Alzar and P. Hyllus for useful discussions.


\begin{thebibliography}{99}
\bibitem{nota2} We indicate with $\langle \hat{O} \rangle_{\inp} \equiv \Tr[\hat{O} \hat{\rho}_{\inp}]$ 
($\langle \hat{O} \rangle_{\out} \equiv \Tr[\hat{O} \hat{\rho}_{\out}]$) the 
expectation value of operator $\hat{O}$ calculated on the initial (final) state, input (output) of the interferometer.
The input and output states are related by $\hat{\rho}_{\out}= e^{\theta (\hat{a}^{\dag}\hat{b}-\hat{b}^{\dag}\hat{a})/2} 
\hat{\rho}_{\inp} e^{-\theta (\hat{a}^{\dag}\hat{b}-\hat{b}^{\dag}\hat{a})/2}$.
\bibitem{nota0} The single mode density matrix $\hat{\rho}_a$ containing, in average, 
$\langle \hat{n}_a \rangle \equiv \mathrm{\Tr}[\hat{\rho}_a \hat{a}^{\dag} \hat{a}]$ particles, 
is called ``number squeezed'' if it has a sub-Poissonian particle-number distribution:
$(\Delta \hat{n}_a)^2 \equiv \mathrm{Tr}[\hat{\rho}_a (\hat{a}^{\dag}\hat{a}-\langle \bar{n}_a \rangle)^2] < \langle \hat{n}_a \rangle$.
\bibitem{note_Schwinger} 
The operators 
$\hat J_x=(\hat a^\dagger\hat b+\hat b^\dagger\hat a)/2$,
$\hat J_y=(\hat a^\dagger\hat b-\hat b^\dagger\hat a)/2i$,
and $\hat J_z=(\hat a^\dagger\hat a-\hat b^\dagger\hat b)/2$
are the basis of a SU(2) group.
\bibitem{notaM}  After the $M$th QND interaction the density matrix is given by 
$\hat{\rho}_A^{(M)} = \mathrm{Tr}_L \Big[\hat{\rho}_{AL}^{(M-1)} |p_M\rangle \langle p_M |\Big] / P_M(p_M)$, 
where $\hat{\rho}_A^{(M-1)} = \mathrm{Tr}_L \Big[\hat{\rho}_{AL}^{(M-2)} |p_{M-1}\rangle \langle p_{M-1} |\Big] / P_{M-1}(p_{M-1})$, etc.
\bibitem{notanotice} Notice that for $\gamma=0$, Eq.(\ref{Dtheta}) predicts $\Delta \theta = \sqrt{2}/\sqrt{\langle\hat{n}\rangle}$, 
where the factor $\sqrt{2}$ is due to the fact that the optimal phase shift is $\theta=\pi/2$.
\end{thebibliography}
\end{document}